\documentclass[aps,prl,twocolumn,groupedaddressm,floatfix]{revtex4}
\usepackage{graphicx}

\begin{document}


\title{Phase coexistence in Gallium nanoparticles controlled by electron excitation}


\author{S. Pochon}
\author{K.F. MacDonald}
\author{R.J. Knize}
\email[on sabbatical leave from the US Air Force Academy, Colorado
Springs, USA]{}
\author{N.I. Zheludev}
\email[]{n.i.zheludev@soton.ac.uk}
\affiliation{School of Physics and Astronomy, University of
Southampton, SO17 1BJ, UK}



\begin{abstract}
 In Gallium nanoparticles of 100 nm in diameter grown on the tip of an optical fiber from an atomic beam we
 observed equilibrium coexistence of $\gamma$, $\beta$ and liquid structural phases that
 can be controlled by e-beam excitation in a highly reversible and reproducible fashion.
 With 2 keV electrons only 1 pJ of excitation energy per nanoparticle  is needed to exercise control,
 with the equilibrium phase achieved in less than a few tenths of a microsecond. The transformations between
 coexisting phases are accompanied by a continuous change in the nanoparticle
 film's reflectivity.
\end{abstract}


\maketitle

Electron beams provides a very fine tool to study small particles,
not only for imaging, but also for preparing excited states of
matter. For instance, delicate stimulation under an electron
microscope has allowed the observation of structural instabilities
in very small metallic clusters \cite{Williams}, \cite{Iijima} and
revealed the complexity of nanoparticle plasmonic excitation
stimulated by the electron beam \cite{Yamamoto}. Our study of
electron beam excitation of gallium nanoparticles is motivated by
the desire to understand the exciting physics of phase equilibria
in nanoparticles \cite{Smirnov94-PS50, Berry84-PRA30,
Berry00-JCP113} and in particular in metallic nanoparticles, which
have the potential to play the key role in future highly
integrated photonic devices as the active elements of waveguiding
\cite{JRKrennPhysRevLett} and switching
\cite{NIZheludevContPhys43} structures. Here we report that
\emph{controllable}, \emph{continuous} and \emph{reversible} phase
coexistence of different crystalline and disordered phases can be
achieved in gallium nanoparticles under electron-beam excitation.

Excitation of gallium nanoparticles with an electron beam is a
multi-stage process, resulting in heating through the loss of
kinetic energy of bombarding electrons and excitation of the
electronic sub-structure of gallium which has elements of covalent
bonding. The energy of the 2 keV electrons used in our experiments
is not sufficient to damage the nanoparticles' material by direct
displacement of gallium atoms, but it is above the 2p and 2s
electron removal thresholds and is therefore sufficient for
multiple ionization of Ga atoms by electron impact: $e + Ga
\longrightarrow e + Ga^{n+} + ne$ up to $n$ = 4. The single
electron ionization cross-section is about $\sigma_1 \approx
0.7\times10^{-16} cm^2$ while the total higher order contribution
is about $\sigma_{2+3+4} \approx 0.2\times10^{-16} cm^2$
\cite{CJPatton}. 2 keV electrons provide relatively even
excitation of the nanoparticle volume as the first ionization
absorption depth in solid Ga is about 50 nm:
$r={2.76\times10^{-2}AE_0^{1.67}/\rho Z^{0.89}}$
where $A = $ 69.723 is the atomic mass, $E_0$ the accelerating
voltage (keV), $\rho$ = 5.91 $gcm^{-3}$ the density, and Z = 31
the atomic number of the gallium target \cite{KanayaJPhysD}. The
resultant secondary electrons and holes created in the Auger
process cause further ionization, generating an avalanche of
electrons which develops and decays on the sub-picosecond time
scale, from the initial electron impact, creating heat and
high-density electrons, hall-pairs and plasmon excitations which
can affect the phase equilibrium of the nanoparticle.

As a playfield to study phase equilibria gallium is a unique metal
in that ten structural solid phases are known. Five phases
$(\alpha, \beta, \gamma, \delta, \varepsilon)$ can exists at low
pressure \cite{ADefrain}. We studied electron-beam induced
structural transformations in Ga nanoparticles on the tip of a
silica optical fiber. The stimulated structural transformations
were detected optically by monitoring the nanoparticle film
reflectivity through the fiber. Due to the very significant
differences in the electronic and optical properties of the
various phases of gallium \cite{Bernasconi95-PRB52}, optical
measurements provide a very sensitive tool for detecting
nanoparticle phase composition.

A schematic of the apparatus is shown in Fig. 1. Nanoparticles
were grown using the recently developed light-assisted deposition
technique \cite{KFMacDonald}. This technique yielded particles of
relatively narrow size distribution with diameters of about 100
nm. The fiber core (9 $\mu m$ in diameter) contained approximately
$6.10^3$ nanoparticles. All experiments were conducted in a vacuum
chamber evacuated to $10^{-6}$ mbar. In the chamber, the fiber tip
supporting the nanoparticles was attached to the cold-finger of a
nitrogen flow cryostat providing temperature scan capability.
After the experiments, the fiber was removed from the vacuum
chamber and examined with an atomic force microscope.

\begin{figure}[h]
    \includegraphics[width=85mm]{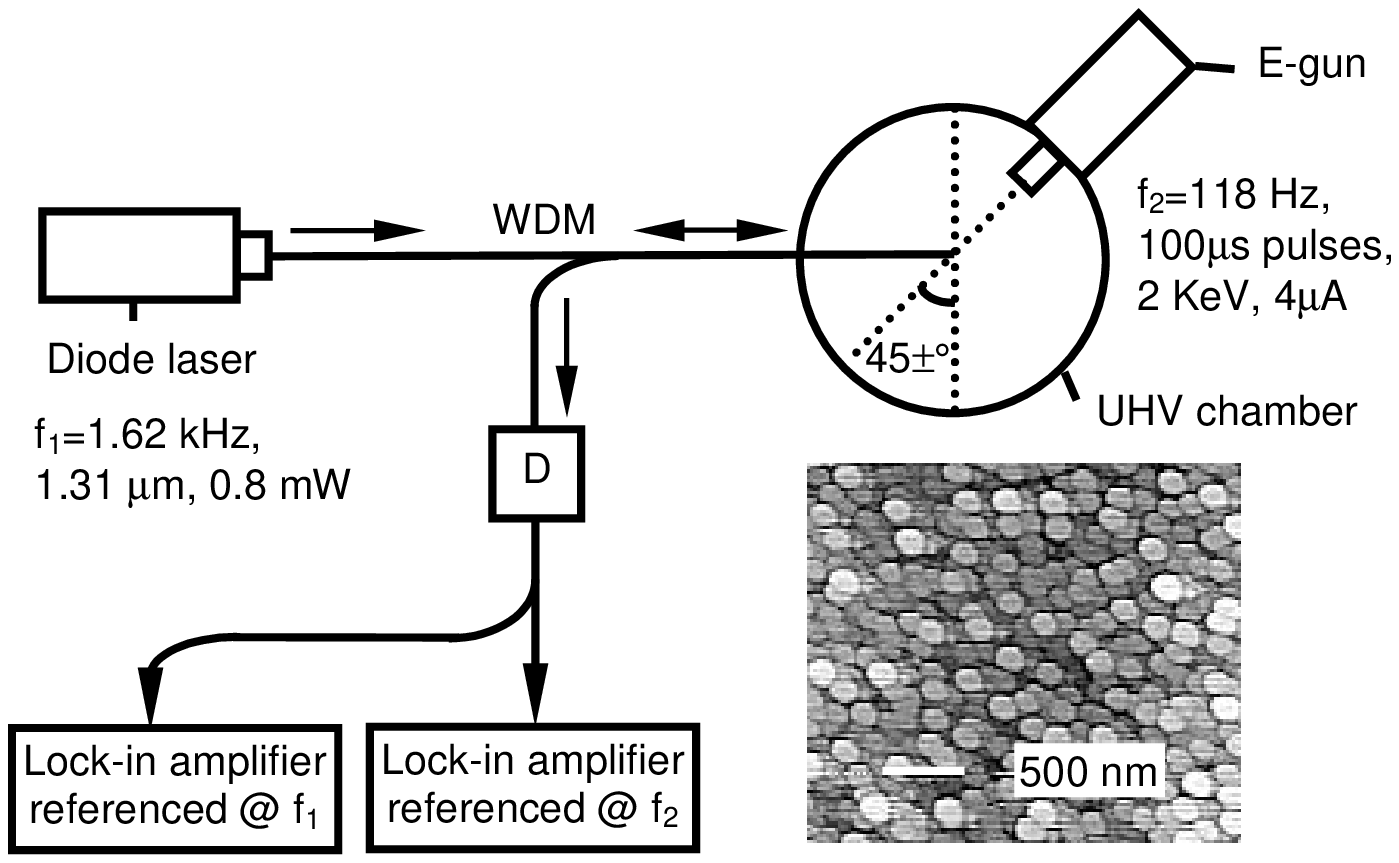}\\
  \caption{Schematic diagram of the apparatus. The inset shows an atomic
    force microscope (AFM) image of gallium nanoparticles located on the core
     area of a fiber (actual size of the particles is a factor of 1.6
     smaller than appears on the picture due to the microscope instrumental function) }\label{setup}
\end{figure}
The nanoparticles were stimulated by a 2 keV electron beam focused
to a spot of about 100 $\mu m$ in diameter to encompass all of the
nanoparticles on the fiber core. The electron gun current of 4
$\mu A$ used in our experiments corresponded to an electron beam
intensity at the fiber core of approximately 100 $Wcm^{-2}$. The
e-beam was modulated to give 100 $\mu$s pulses with repetition
rate of 118 Hz, providing average power of about 120 pW per
nanoparticle, or a total energy of 1pJ per pulse per nanoparticle.
To monitor the film's reflectivity we used a diode laser operating
at 1.31 $\mu m$ with a power of 800 $\mu W$, modulated at a
frequency of 1.62 kHz. The intensity of reflected light was
detected with two phase-sensitive amplifiers. One amplifier was
locked at the frequency of electron beam modulation, to detect
electron-beam induced effects, the other was locked at the probe
beam modulation frequency, to monitor variations of the sample
reflectivity. Our experiments were performed in the temperature
range from 80 to 250 K. The temperature of the coldfinger was
measured with an absolute accuracy better that 0.5 K, but was
somewhat lower than the actual temperature of the nanoparticles at
the fiber tip due to the electron and laser heating. The
temperature scale presented here in the experimental graphs takes
this into account and was calibrated on the melting temperature of
the gallium nanoparticles.

Structural transformations in the nanoparticles were observed by
monitoring the nanoparticle film reflectivity and electron-beam
induced reflectivity changes, recorded during heating-cooling
cycles. Reflectivity recorded during the first heating-cooling
cycle after growth is shown in Fig. 2a (bold curve). The
reflectivity showed an increase during heating at about 120 K and
a much larger increase which begins around 230 K, with a total
reflectivity change of about 1.7 $\%$. The reflectivity remains
high during cooling down to about 145 K, where it rapidly
decreases to form an incomplete hysteresis loop about 100 K wide.
In the next heating-cooling cycle (faint curve), the hysteresis
loop remains very wide, but becomes much more shallow (about
0.75$\%$ of total change) and nearly complete.
\begin{figure}[h]
    \includegraphics*[width=90mm]{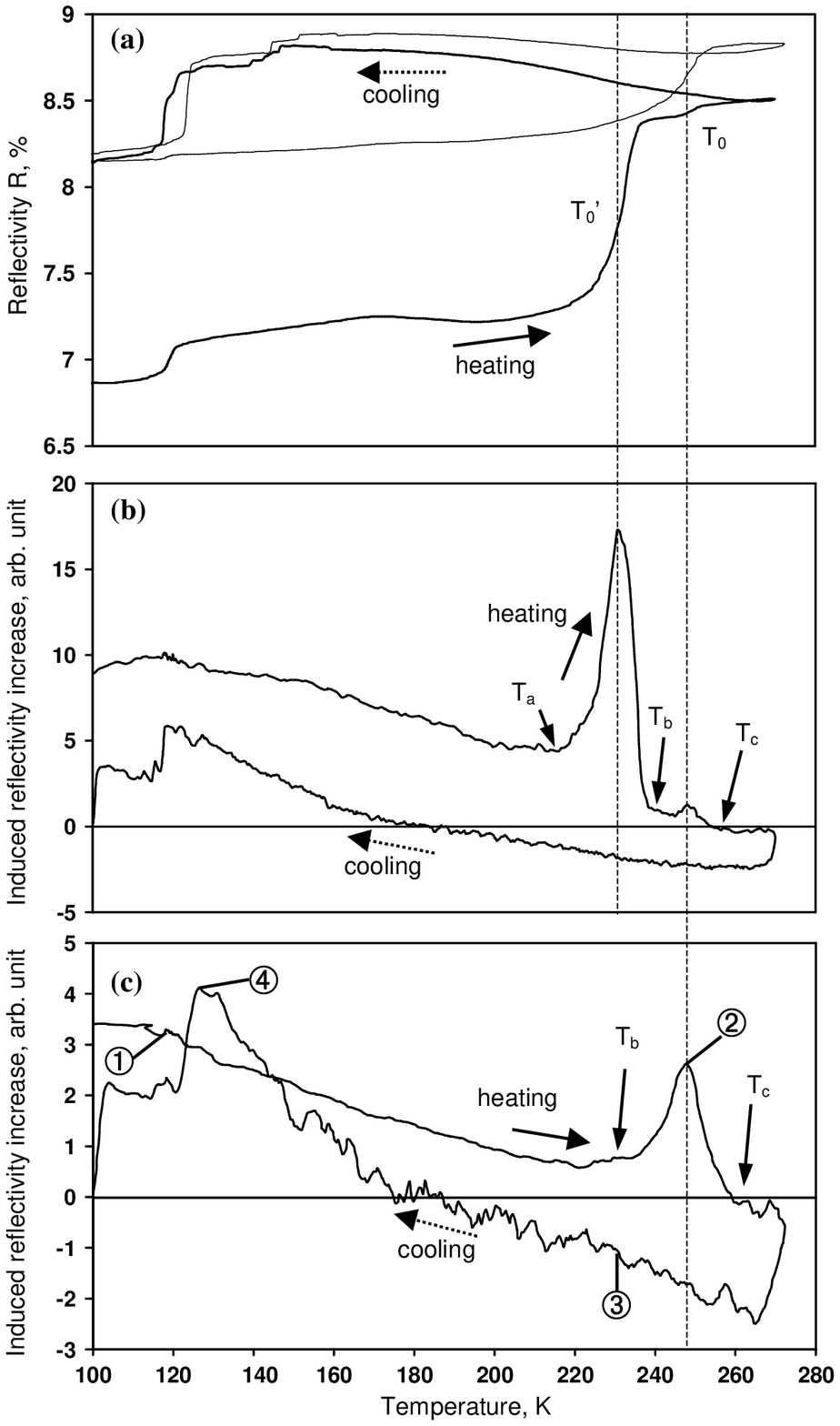}\\
  \caption{Temperature dependences of (a) reflectivity R, (b)
  electron-beam induced reflectivity increase during the first temperature scan, and (c)
  electron-beam induced reflectivity increase for following scans.}
\end{figure}

The modulated electron beam induces an increase in the optical
reflectivity as  shown in Fig. 2b for the first heating-cooling
cycle. A large peak is observed at 231 K and a smaller peak at 248
K. At temperatures above the second peak, the electron-beam
induced signal becomes negative (reflectivity decreases), and
remains negative during cooling down to about 180 K. On the
cooling part of the curve a peak is seen at 120 K, corresponding
to the reflectivity drop. The second and subsequent temperature
scans show that the first rising temperature peak disappears and
the second peak (at 248 K) increases (Fig. 2c). We also performed
measurements of the transient dynamics of the induced reflectivity
change with 2 $\mu$s electron pulses. Reflectivity was then
monitored using an amplified photodetector and real time digital
scope. We observed essentially non-exponential reflectivity
relaxation and an increase in the relaxation time of the response
at temperatures where the nonlinear response is peaking (Fig. 3).

\begin{figure}[h]
    \includegraphics[width=90mm]{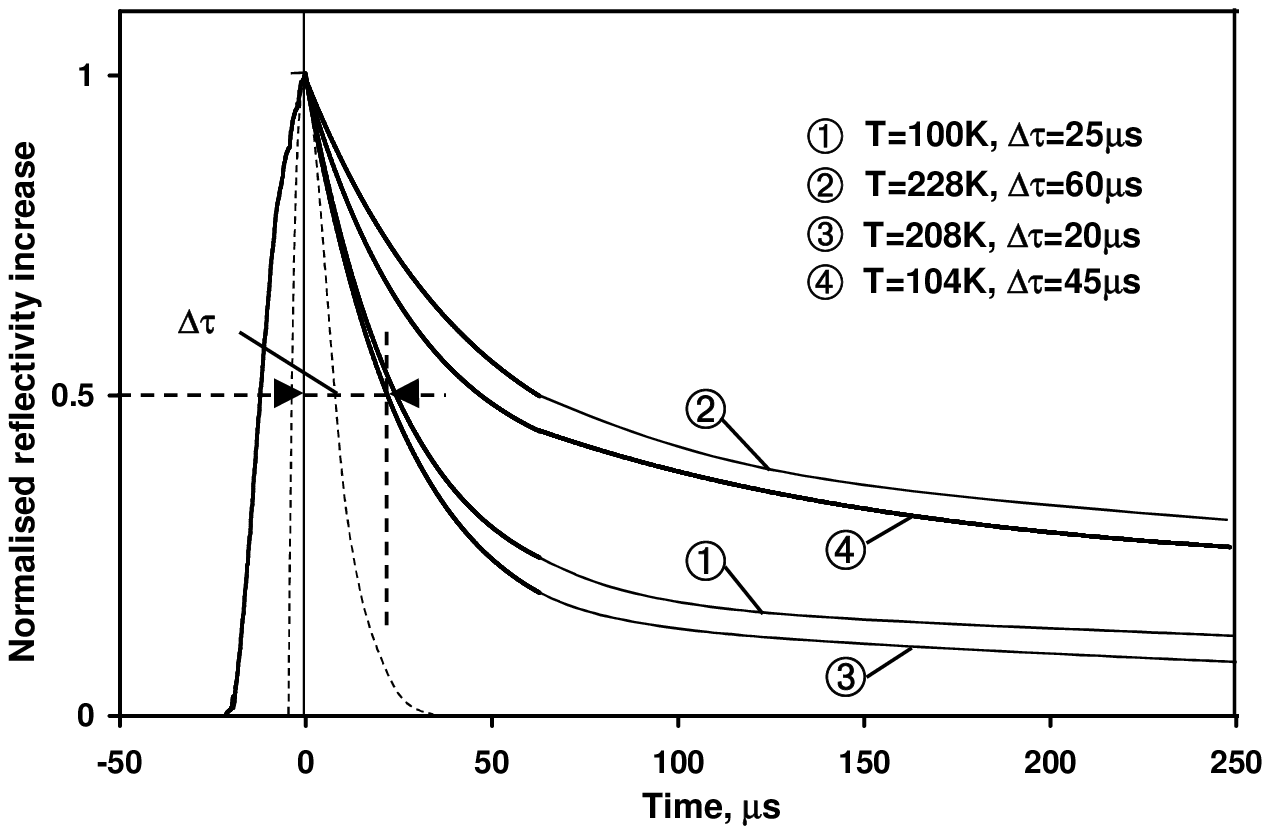}\\
  \caption{Transient reflectivity increase of a gallium nanoparticle
  film to 2 $\mu$s electron beam pulses (normalised) for the four numbered
  positions shown in Fig. 2c. Characteristic relaxation times ($\Delta$$\tau$) are
  measured at half maximum. The doted curve shows
   response of the detector system to a 3 ns optical pulse.}
\end{figure}

We argue that our experiment reveals a reversible electron-beam
induced structural transformation in nanoparticles in the form of
a controlled  dynamic coexistence between different structural
forms. The surface of a particle, where atoms have fewer nearest
neighbors than internal atoms, acts as a boundary at which
transformation processes start. To detail this process  further,
we shall consider a nanoparticle with a core consisting of a
certain structural phase covered by a `shell' of different
composition. Comparison with energy-dispersive x-ray diffraction
studies of gallium nanoparticles \cite{DiCicco98-PRL81} suggests
that the low-temperature phase is $\gamma$-gallium, with a bulk
melting temperature of 238 K, and that the intermediate phase is
$\beta$-gallium, with a bulk melting temperature of 257 K (these
studies indicate that $\alpha$-gallium is not found in
nanoparticles). The melting temperatures in nanoparticles are
depressed in comparison with those of the bulk by $\delta T = K/d$
where $d$ is the nanoparticle diameter. For the $\beta$ phase K =
600 K.nm \cite{EVCharnaya}. This gives the $\beta$-gallium
nanoparticles a melting temperature of about 251 K. In the
simplest case of a phase transition to the melt in the
nanoparticle, the electron-beam induced behavior should be
analogous to the temperature-driven `surface melting' effect that
has been seen in lead nanoparticles \cite{Garrigos89-ZPD12} and
found to be thermodynamically reversible within a narrow
temperature range \cite{Peters97-APL71}. In reality the situation
is complicated by the presence of two steps in the first
reflectivity dependence (Fig. 2a) at $T_0$$^{'}$ and $T_0$ (bold
curve). It is instructive to note that assuming the same
depression coefficient for all phases involved, the difference
between the melting points of the $\gamma$ and $\beta$ phases is
about 19 K which is close to the 17 K difference observed between
the first and second peaks on Fig. 2b, indicating that $T_0$$^{'}$
is the melting temperature of the $\gamma$ phase, and that it is
the ground state phase of the nanoparticle after fabrication. In a
multi-phase nanoparticle there are two possible scenarios, either
the nanoparticles first undergo a transition from one solid phase
to another at $T_0$$^{'}$, and then from that phase to the liquid
at $T_0$, or different solid phases with different melting points
initially coexist in nanoparticles at low temperature. In the
presence of electron excitation, the phase equilibrium will be
determined by both temperature and electron-beam intensity.

In the first transition scenario the influence of excitation on
the equilibrium becomes apparent in the changing reflectivity of
the film at a temperature $T_a$ below $T_0$$^{'}$ (see Fig. 2b).
With increasing temperature or level of excitation, the
$\beta$-gallium surface layers' thickness increases until the
transformation of the $\gamma$-gallium core to the `surface' phase
is completed. When, at $T_b$, the core of the particle is fully
consumed by the $\beta$ phase the nanoparticle becomes stable
against a return to the $\gamma$ phase because this would require
the creation of a nucleation center. However, if the temperature
or level of electronic excitation is reduced \emph{before} the
transformation to the new phase is complete, i.e. while a nucleus
of the old core phase is still present, the transformation is
reversed and the skin layer shrinks to an appropriate equilibrium
position. Thus, reversibility is provided in the temperature range
between $T_a$ and $T_b$. This whole process is then replicated
between $T_b$ and $T_c$, the $\beta$ and liquid phases around the
next transition temperature $T_0$, (see Fig. 4a). It then appears
that, on cooling the nanoparticles return to the $\beta$ phase but
the $\gamma$- phase is not present anymore. This is evident from
the shallower reflectivity hysteresis. During the second and
following temperature cycles the nanoparticles in the $\beta$
phase only go through the second stage of transformation, as
presented in Fig. 4c.

In the second scenario, the $\gamma$ and $\beta$ phases coexist in
gallium nanoparticles after their formation on the substrate from
the atomic beam. The first temperature cycle then shows
consecutive melting of the $\gamma$ component at $T_0$$^{'}$ and
of the $\beta$ component at $T_0$, as presented in Fig. 4b. It is
not, however, possible to distinguish between the two
transformation scenarios outlined above on the evidence of the
reflectivity data available to us and this shall be left for
further investigations. It is also possible that the two-step
process observed during the first temperature cycle results from
the coexistence of nanoparticles of different ground states
immediately after growth. Whatever the phase composition of the
particles, the longer relaxation observed around the peaks in the
nonlinear response indicate an increase in the time needed for the
phase boundary to travel across the increasingly thick 'shell'
layer.

\begin{figure}[!h]
\includegraphics[scale=0.8]{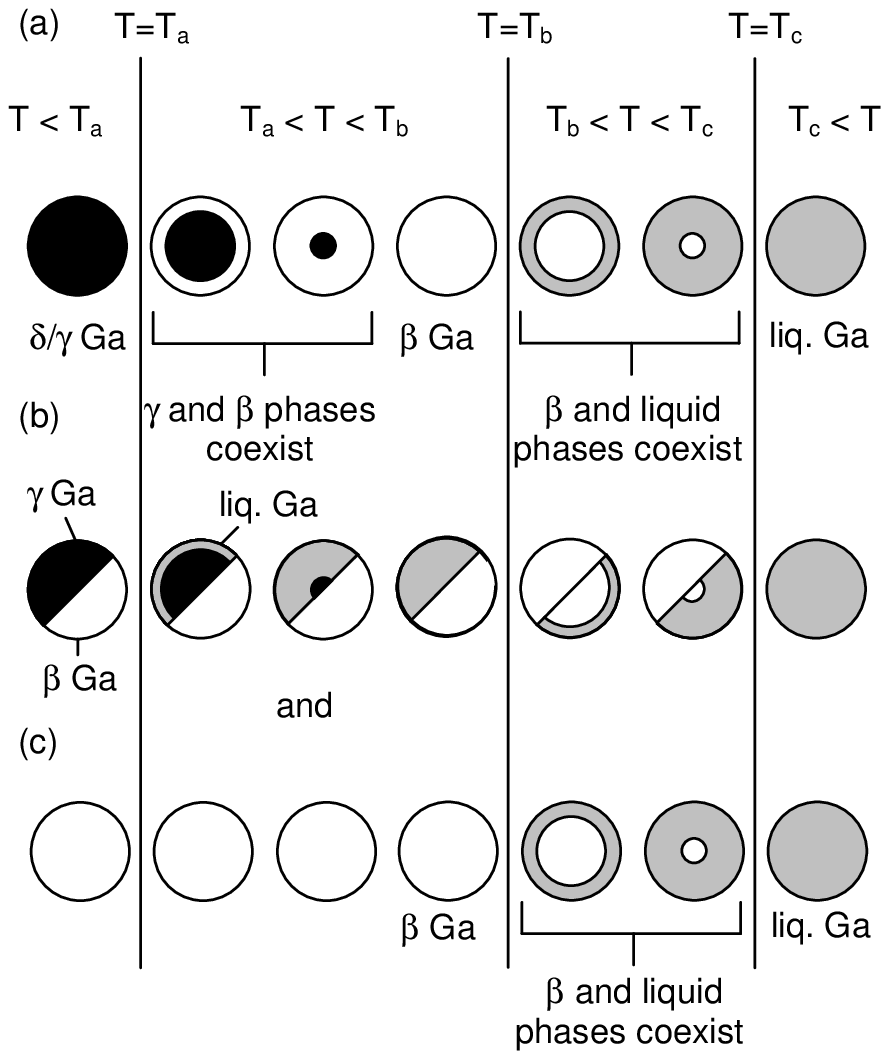}
\caption{Schematic representation of the phase coexistence
scenarios in gallium nanoparticles. (a) solid-solid-liquid
transition (first temperature cycle), (b) coexistence of different
solid phases and two overlapping solid-liquid transitions
(alternative scenario for the first temperature cycle), and (c)
Solid-liquid transition (following temperature cycles).}
\end{figure}

The strength of the phase coexistence concept is supported by our
calculations of the optical properties of gallium nanoparticle
films on a dielectric substrate using a recently developed
effective-medium model for densely packaged nanoshells
\cite{JOpt}. For the purposes of our calculations, the dielectric
constants of $\beta-$ and $\gamma-$ gallium, which are much closer
to those of a free-electron metal than those of the $\alpha$
phase, were estimated by using the damping constant in Drude's
free-electron model as a fitting parameter to produce the
nanoparticle film reflectivity levels shown in Fig. 2a. These
calculations confirmed that the presence on each nanoparticle of a
shell just a few nanometres thick in a phase different from the
core can produce a change in reflectivity sufficient to explain
our experimental data.

A thermally activated transition due to electron-beam-induced
heating can explain certain characteristics of the effect. For
instance, by assuming a local electron-induced temperature
increase of $4$ $K$, one can derive a good facsimile of the
experimental peaks in induced reflectivity increase at $T_0^{'}$
and $T_0$ from the reflectivity data in Fig. 2a. However, there
are serious discrepancies between the results of this thermal
model and the experimental results, primarily at temperatures more
than a few degrees below the peaks, where the observed effect is
larger than predicted by the thermal model. This suggests that
another, temperature-independent, non-thermal excitation mechanism
is also contributing to the effect. This mechanism may be
especially important for $\beta$-gallium because its structure
contains covalent bonds \cite{Bernasconi95-PRB52}. As with the
excitation mechanism in e-beam pumped semiconductor lasers,
electron-beam excitation in gallium results in bonding-antibonding
transitions, which destabilize the crystalline structure
\cite{Siegal95-ARMS25}. This mechanism should be especially
effective in nanoparticles as the electron-beam penetration depth
in gallium is of the order of their diameter. An `inclusion' of a
new phase is thus created, changing the optical properties of the
`host' phase at temperatures far below its transition point and
shifts the phase equilibrium, promoting the formation of a thicker
layer of the new phase without any increase in temperature
\cite{MacDonald01-JOSAB18}.

In conclusion, we observed equilibrium coexistence of different
structural phases in gallium nanoparticles that can be controlled
by e-beam excitation in a highly reversible and reproducible
fashion.

The authors acknowledge the financial support of the EPSRC (UK).

\end{document}